\documentstyle[12pt,twoside]{article}
\def\greaterthansquiggle{\raise.3ex\hbox{$>$\kern-.75em\lower1ex\hbox{$\sim$}}}
\def\lessthansquiggle{\raise.3ex\hbox{$<$\kern-.75em\lower1ex\hbox{$\sim$}}}
\newcommand{\beq}{\begin{equation}}
\newcommand{\eeq}{\end{equation}}
\newcommand{\beqa}{\begin{eqnarray}}
\newcommand{\eeqa}{\end{eqnarray}}
\newcommand{\beqan}{\begin{eqnarray*}}
\newcommand{\eeqan}{\end{eqnarray*}}
\newcommand{\ba}{\begin{array}}
\newcommand{\ea}{\end{array}}

\newcommand{\ra}{\rightarrow}

\newcommand{\ve}{\varepsilon}
\newcommand{\vp}{\varphi}

\newcommand{\wt}{\widetilde}

\newcommand{\R}{{\cal R}}

\newcommand{\st}{\stackrel}

\def\nz{\ifmmode {I\hskip -3pt N} \else {\hbox {$I\hskip -3pt N$}}\fi}
\def\zz{\ifmmode {Z\hskip -4.8pt Z} \else
       {\hbox {$Z\hskip -4.8pt Z$}}\fi}
\def\qz{\ifmmode {Q\hskip -5.0pt\vrule height6.0pt depth 0pt
       \hskip 6pt} \else {\hbox
       {$Q\hskip -5.0pt\vrule height6.0pt depth 0pt\hskip 6pt$}}\fi}
\def\rz{\ifmmode {I\hskip -3pt R} \else {\hbox {$I\hskip -3pt R$}}\fi}
\def\cz{\ifmmode {C\hskip -4.8pt\vrule height5.8pt\hskip 6.3pt} \else
       {\hbox {$C\hskip -4.8pt\vrule height5.8pt\hskip 6.3pt$}}\fi}

\normalsize
\begin{document}
\bibliographystyle{plain}
\begin{titlepage}
\begin{flushright}
UWThPh-1995-24 \\
\today
\end{flushright}
\vspace{1cm}
\begin{center}
{\Large \bf
Vacuum Spacetimes with Future Trapped Surfaces}\\[1 cm]
R. Beig*  \\
Institut f\"ur Theoretische Physik \\
Universit\"at Wien \\
Boltzmanngasse 5, A--1090 Wien, Austria\\
Fax: ++43-1-317-22-20, E-mail: BEIG@PAP.UNIVIE.AC.AT \\[4pt]
and \\[4pt]
N. \'O Murchadha** \\
Physics Department \\
University College Cork \\
Cork, Ireland\\
Fax: +353-21-276949, E-mail: NIALL@BUREAU.UCC.IE
\vfill
{\bf Abstract} \\
In this article we show that one can construct initial data for the Einstein
equations which satisfy the vacuum constraints. This initial data is defined on
a manifold with topology $R^3$ with a regular center and is asymptotically
flat. Further, this initial data will contain an annular region which is
foliated by two-surfaces of topology $S^2$. These two-surfaces are future
trapped in the language of Penrose. The Penrose singularity theorem guarantees
that the vacuum spacetime which evolves from this initial data is future null
incomplete.
\end{center}

\vfill
\noindent *) Supported by Fonds zur F\"orderung der wissenschaftlichen
Forschung in \"Osterreich, Project No. P9376--PHY. \\{}
**) Partially supported by Forbairt Grant SC/94/225.
\end{titlepage}

\section{Introduction}
\renewcommand{\theequation}{\arabic{section}.\arabic{equation}}
\setcounter{equation}{0}

This paper is the third in a series dealing with the existence of globally
regular, asymptotically flat, maximal initial data for General Relativity
in vacuo which have trapped surfaces. Our approach consists of
considering certain classes of sequences of initial data called critical
sequences (CS's) which, for large values of the parameter, develop
regions closely resembling the black hole region of the extended
Schwarzschild (SS--) spacetime. We first recall (see [1,2]) the properties
of the SS--geometry which form the model for our more general results.

Consider the spacetime metric
\beq
ds^2 = -\left(1 - \frac{2m}{\wt r}\right) dt^2 + \left(1 - \frac{2m}{\wt r}
\right)^{-1} d \wt r^2 + \wt r^2 dO^2
\eeq
where $0 < 2m < \wt r < \infty$, $- \infty < t < \infty$ and $dO^2$ is
the line element on the unit two--sphere.
The spacetime (1.1) can be viewed as the right quadrant of the Kruskal
manifold. In particular it can be smoothly extended to the black hole
region (i.e. the upper quadrant of Kruskal) by replacing $t$ by the
Eddington--Finkelstein coordinate
$\tau = t + \wt r + 2m \log |\frac{\wt r}{2m} - 1|$ and allowing $\wt r$ to
vary over
$0 < \wt r < \infty$ and $- \infty < \tau < \infty$. For $\wt r < 2m$
we can, if we wish, again represent $ds^2$ by Equ. (1.1). Now fix a
constant $C > 0$ with
\beq
C^2 < \frac{27m^4}{16}
\eeq
and consider the function $\vp : {\bf R}^+ \ra {\bf R}$ defined by
\beq
\vp(\wt r) = 1 - \frac{2m}{\wt r} + \frac{C^2}{\wt r^4}.
\eeq
$\vp$ is negative for $\wt r = 3m/2$, positive for $\wt r = 2m$, goes to
one as $\wt r \ra \infty$ and is monotonically increasing for
$\wt r \geq 3m/2$. Thus, for some unique $\wt r_c$ with
$3m/2 < \wt r_c < 2m$, $\vp(\wt r_c) = 0$ and $\vp > 0$ for $\wt r > \wt r_c$.
Moreover $\wt r_c$ is a simple zero of $\vp$. Now define a function $f$
$(\wt r \geq \wt r_c)$ by
\beq
f(\wt r) = -C \int_{r_c}^{\wt r} \frac{ds}{s^2 (1 - 2m/s)
\sqrt{\vp(s)}}.
\eeq
This will be the height-function of a spherically symmetric maximal slice
through the Schwarzschild solution of the kind discussed in [1,2].
If (1.4) is taken in the sense of the Cauchy mean value at $s = 2m$,
$f(\wt r)$ is defined for $\wt r_c \leq \wt r < 2m$ and
$2m < \wt r < \infty$. One can see that $f(\wt r)$ is everywhere positive and
is logarithmatically divergent at $\wt r = 2m$. One can also see that
the function $g$ given by
\beq
g(\wt r) = f(\wt r) + 2m \log |\frac{\wt r}{2m} - 1| + \wt r
\eeq
is $C^\infty$ for all $\wt r \in (\wt r_c,\infty)$. Thus the height-function
diverges logarithmically at the horizon in the standard Schwarzschild
coordinates but is everywhere regular in the regular Eddington-Finkelstein
coordinates. Now consider the subset (of the extended manifold) given by
$\tau = g(\wt r)$. For each fixed $C$ satisfying (1.2) this is a smooth
submanifold along which, for $\wt r \neq 2m$, one has  $t = f(\wt r)$. This
will lie in the  upper half of the right-hand quadrant since we know that $f >
0$ and continues  through to the middle of the upper quadrant of the Kruskal
diagram.  We can express this as
$\wt r = f^{-1}(t)$, $t > 0$. The latter function can be smoothly
extended to negative values of $t$ by writing $\wt r = f^{-1}(|t|)$.
This results in an extended submanifold $\wt M_C$ which is a Cauchy slice
of the Kruskal manifold and lies symmetrical with respect to the
timelike cylinder given by $t = 0$. (Note that the set where $t = 0$
consists of a time symmetric Cauchy slice plus a timelike cylinder
which intersect orthogonally along the bifurcation sphere of the
horizon.) The two asymptotic ends of $\wt M_C$ are translates under
$\partial/\partial t$ of one another by the amount
$\Delta t = 2 \lim_{\wt r \ra \infty} f(\wt r)$. This surface $\wt M_C$
lies is the upper half of the Kruskal diagram. The `mirror' surface in the
lower half plane corresponds to choosing $C < 0$.

One can now compute the first and second fundamental form of $\wt M_C$.
One gets $(\wt r_c < r < \infty)$
\beq
\wt g_{ab} dx^a dx^b = \frac{1}{\vp} d \wt r^2 + \wt r^2 dO^2
\eeq
and
\beq
\wt p_{ab} dx^a dx^b = C \left( \frac{2}{\vp \wt r^3} d \wt r^2 -
\frac{1}{\wt r} dO^2 \right),
\eeq
where $a$, $b$ run from 1 to 3. Note that our choice of the definition of
extrinsic curvature starts from $K_{\mu\nu} = +{1 \over 2}{\cal L}_
u\gamma_{\mu\nu}$ which agrees with Wald[3] but is the negative of the choice
made in MTW[4]. It now follows that $\wt p = \wt p_{ab} \wt g^{ab} = 0$ and
that $\wt p^{ab}_{\ \ ;a} = 0$. There is a unique (up to a constant factor)
spherically symmetric 2-tensor which is both divergence-free and trace-free
(TT-) on flat space [5]. TT-tensors are conformally covariant. The metric $\wt
g_{ab}$ above is conformally flat and the tensor $\wt p_{ab}$ is the
conformal transform of the flat-space spherically symmetric TT-tensor. In flat
space the TT-tensor blows up at the origin, but since the  slice through
Schwarzschild has topology $S^2 \times {\bf R}$ it is everywhere regular.

The complete 3--manifold
$\wt M_C$ consists of two copies of the asymptotically flat manifold (1.6,7),
glued together along the minimal surface $\wt r = \wt r_c$ where
$\vp$ vanishes. (The subsets $\wt r > 2m$ of these two copies are
contained in the right and left quadrant of Kruskal, respectively.)
Thus the
$\wt M_C$'s are maximal spherically symmetric Cauchy surfaces. Since $\wt r_c >
3m/2$, they stay away from the singularity $r = 0$ uniformly in $C$. It is well
known that the spheres of constant $t = t_0$, $\wt r = \wt r_0$ for
$\wt r_0 < 2m$ are future trapped surfaces (FTS's). For spheres
lying in $\wt M_C$ for some $C$, we can express this fact in terms of
$\wt g_{ab}$ and $\wt p_{ab}$. The divergence of the null normals to a
submanifold $\Sigma$ of a spacelike slice $\wt M$ is given in general
by (note that our choice of sign for $\wt p_{ab}$ plays a role here)
\beq
\Theta_\pm = H \mp \wt p_{ab} \wt n^a \wt n^b \pm \wt p,
\eeq
where + $(-)$ refers to the future (past) outer null normal,
$\wt n^a$ is the outer normal to $\Sigma$ embedded in $\wt M$ and $H$
the mean curvature of $\Sigma$ w.r. to $\wt n^a$. ``Outer'', in the
present context, is just conventional, say ``right''. The surface
$\Sigma$ is called future--trapped iff $\Theta_+ < 0$ and $\Theta_- > 0$.
In such a situation the Penrose singularity theorem [6] implies that
any vacuum spacetime Cauchy evolving from $M$ is future null incomplete.

Suppose, contrary to the Schwarzschild situation, that $\Sigma$ divides $M$
into an outer, non--compact region (containing spatial infinity) and a
compact inner region. In this case
$\sigma$ is called an outer trapped surface (OTS) when merely
$\Theta_+ < 0$. (Note that the notion of OTS, as opposed to that of FTS,
is not necessarily time--asymmetric. We could have surfaces $\Sigma$
for which both $\Theta_+$ and $\Theta_-$ are negative.) It is known
(see [7]) that the OTS--condition is already sufficient for the Penrose
theorem to hold.

Let us remark at this point that the time--symmetric slices of SS
have neither FTS's nor OTS's (at least no spherically symmetric ones),
so the Penrose singularity theorem does not apply. But, due to the
existence of two infinities (since the topology is $S^2 \times {\bf R})$,
the singularity theorems of Gannon [8] and Lee [9] can be used,
again with the result that any Cauchy evolution is geodesically incomplete.

Applying (1.8) to $\wt r =$ const in $M_C$ we find that
\beq
\Theta_\pm = \frac{2}{\wt r} \left( \sqrt{\vp} \mp \frac{C}{\wt r^2}
\right)
\eeq
for the right copy and
\beq
\Theta_\pm = \frac{2}{\wt r} \left( - \sqrt{\vp} \mp \frac{C}{\wt r^2}
\right)
\eeq
for the left copy. Since $\vp > C^2/\wt r^4$ for $\wt r > 2m$ and
$\vp < C^2/\wt r^4$ for $\wt r < 2m$, $\Theta_+$ and $\Theta_-$ are both
positive for $\wt r > 2m$ in the right copy. At $\wt r = 2m$ the quantity
$\Theta_+$ changes sign and $\Theta_-$ remains positive. As we cross
$\wt r = 2m$ in the left copy, $\Theta_+$ remains negative and $\Theta_-$
changes sign. In particular, all spheres of constant $\wt r_0$ in
$\wt M_C$ with $\wt r_0 < 2m$ are FTS's.

All spherically symmetric, complete maximal spacelike slices of SS are
either the $\wt M_C$'s and their translates under $\partial/\partial t$
or the time symmetric slices $t =$ const which can be viewed as limits
of the $\wt M_C$'s as $C$ goes to zero. On the latter slices one has
$\Theta_+ = \Theta_-$ positive on the right copy, zero on the bifurcation
2--sphere and negative on the left copy.

In previous work [10,11] we have constructed sequences of solutions
$(\st{n}{\wt g}_{ab}, \st{n}{\wt p}_{ab})$ with $\wt p = 0$ of the
Einstein vacuum constraints on ${\bf R}^3$ which are asymptotically flat,
complete and which, as $n \ra \infty$, develop annular regions $A_n$
foliated by 2--spheres for which $\Theta_+$ is negative. Although these
data are not necessarily time--symmetric their behaviour in $A_n$ is
dominated by $\st{n}{\wt g}$ for large $n$ and, in particular,
$\Theta_-$ will also be negative in $A_n$. In fact, for large $n$, the geometry
of $A_n$ is approximated by an annulus
of the left (in keeping with the previous convention) copy of a
$t = 0$ slice of SS. For SS of course, the left and right copy
are completely equivalent, but in the CS's we consider the interior of
$A_n$ is topologically a ball rather than a second asymptotic end.
Thus the (extended) Penrose theorem on OTS's applies to spacetimes
evolving from these data, and one concludes that, for large $n$, they
are null geodesically incomplete both in the future and in the past.
In the present work we perform a similar construction, but where we are
able to control the sign of the second term in Equ. (1.8) in such
a way that, for large $n$, we obtain annular regions $A_n$ in our
initial data sets which are modelled on the manifolds
$(\wt M_C, \wt g_{ab},\wt p_{ab})$ of SS described above in an annular
region close to the minimal surface $\wt r = \wt r_c$. In particular
the $A_n$'s for large $n$ are foliated by future trapped surfaces.
Whether, as is suggested by the Schwarzschild case, there is inside
of $A_n$ a region which is both outer future-- and past trapped, is
at present beyond our control. Ultimately we are interested in
initial data with FTS's which result from Cauchy evolution of ones
leaving neither OTS's nor FTS's, or, more ambitiously, which are
asymptotically flat at past null infinity so that they, provided weak
cosmic censorship is true, describe gravitational radiation
collapsing to a black hole. To construct such data, or perhaps to show
that the ones constructed here have this property, is an open problem.

In our most recent work on this topic  we
considered initial data sets with non-zero extrinsic curvature. However, we
assumed that the extrinsic curvature fell off rapidly at infinity and played no
significant role in trapped surface formation. The key improvement in this
current article is that we control the asymptotic behaviour of the extrinsic
curvature. In particular, we assume that the extrinsic curvature near infinity
is dominated by the spherically symmetric TT-tensor of Equ.(1.7). Thus we
construct a family of initial data, in which, to leading order, the intrinsic
geometry looks like (1.6) and the extrinsic curvature looks like (1.7) with
some constant $C$. This means that, to leading order, the null expansions
approximate Equ.(1.9). Further, as in our previous work, we can allow the ADM
mass to become unboundedly large while controlling the error terms. This means
that we can construct initial data for which the now approximate formulae
Equ.(1.9) are accurate for $\wt r \geq 2m$. This initial data set will contain
an annular region around $\wt r = 2m$ of future trapped surfaces with a
minimal surface in the annulus.

\section{Definitions and results from previous work}
\renewcommand{\theequation}{\arabic{section}.\arabic{equation}}
\setcounter{equation}{0}

Let $M$ be a compact manifold diffeomorphic to $S^3$. For $g$ a smooth
Riemannian metric on $M$, define the conformal Laplacian acting on
functions by
\beq
L_g = - \Delta_g + \frac{1}{8} \R[g],
\eeq
where $\Delta_g = g^{ab} D_a D_b$, $D_a$ the covariant derivative and
$\R$ the scalar curvature of $g$. The elliptic operator $L_g$, viewed
as a densely defined operator on the Hilbert space $L^2(M,g)$ is
essentially self--adjoint, with real eigenvalues bounded from below.
Let $\lambda_1(g)$ be it's lowest eigenvalue. Suppose
$\lambda_1(g) > 0$. Let $\Lambda$ be some arbitrary point in $M$.
Then $L_g$ has a unique, positive Green function with source point
$\Lambda$. More precisely, there exists $G : M \setminus \Lambda \ra
{\bf R}$, $G > 0$, satisfying
\beq
\int_M (L_g f)G dV_g = \left. 4 \pi f \right|_\Lambda
\eeq
for all $f \in C^\infty(M)$, or equivalently
\beq
L_g G = \left. 4 \pi \delta \right|_\Lambda,
\eeq
where $dV$ is the Riemannian volume element and
$\left. \delta \right|_\Lambda$ the Dirac delta distribution with source
point $\Lambda$. The singularity of $G$ near $\Lambda$ can be described
as follows. Let $\Omega$ be an asymptotic distance function (ADF).
This should mean that $\Omega > 0$ outside $\Lambda$ and
\beq
\left. \Omega \right|_\Lambda = 0, \qquad
\left. D_a \Omega \right|_\Lambda = 0, \qquad
\left. (D_a D_b \Omega - 2g_{ab})\right|_\Lambda = 0,
\eeq
\beq
\left. D_a D_b D_c \Omega \right|_\Lambda = 0 .
\eeq
This, in a local coordinate neighbourhood centered at $\Lambda$, is
equivalent to
\beq
\Omega = \Omega_0 + \frac{1}{2} g_{ab,c}(0) x^a x^b x^c +
O^\infty(\Omega_0^2),
\eeq
where $\Omega_0 = g_{ab}(0) x^a x^b$ and $f = O^\infty(\Omega_0^{k/2})$
mean $f = O(\Omega_0^{k/2})$, $D f = O(\Omega_0^{(k-1)/2})$,
$DD f = O(\Omega_0^{(k-2)/2})$, a.s.o. In particular
$\Omega = O^\infty(\Omega_0)$. A function of the form
$\Omega = \Omega_0 + O^\infty(\Omega_0^{3/2})$ satisfies (2.4)
but not necessarily (2.5). Given an ADF, $G$ has the property that
\beq
G = \Omega^{-1/2} + \frac{m}{2} + O(\Omega^{1/2})
\eeq
for a constant $m$ and
\beq
\partial (G - \Omega^{-1/2}) = O(1).
\eeq
The constant $m$ has the interpretation of the ADM mass of the metric
$\wt g_{ab} = G^4 g_{ab}$ on $M \setminus \Lambda \cong {\bf R}^3$.
Let $\omega$ be positive. Then
\beq
L_{\bar g} \bar \Phi = \omega^{-5} L_g \Phi
\eeq
for all $\Phi$, where
\beq
\bar g = \omega^4 g, \qquad \bar \Phi = \omega^{-1} \Phi.
\eeq
Letting $\omega = \Phi$ and $\Phi = G$, so that $\bar \Phi \equiv 1$,
it follows from Equ. (2.3) that
\beq
\R[\wt g] = 0 \qquad \mbox{on } \wt M = M \setminus \Lambda
\eeq
where $\wt g_{ab} = G^4 g_{ab}$. By virtue of (2.4,5) and (2.7)
$\wt g_{ab}$ is
asymptotically flat near $\Lambda$. Thus, by the positive--mass theorem
[12,13], we have that $m \geq 0$ and $m = 0$ implies that $\wt g$ is
the flat metric on $\wt M = {\bf R}^3$, or, equivalently, $g$ is
conformal to the standard metric on $S^3$.

Next let $\rho$ be smooth in $M \setminus \Lambda$, the singularity at
$\Lambda$ being restricted by
\beq
\rho = O(\Omega_0^{-3}), \qquad
\partial \rho = O(\Omega_0^{-7/2})
\eeq
and consider the equation
\beq
L_g  \Phi = \left. 4 \pi \delta \right|_\Lambda + \frac{1}{8}
\Phi^{-7} \rho .
\eeq
This is treated as follows. Write
\beq
\Phi = G + h,
\eeq
so that (2.13) gets replaced by
\beq
L_g h = \frac{1}{8} (1 + G^{-1} h)^{-7} G^{-7} \rho.
\eeq
In order to solve Equ. (2.15) we use the Green function $G(x,x')$ with
source at an arbitrary point $x' \in M$. Define
\beq
\Omega_0(x,x') = g_{ab}(x')(x-x')^a(x-x')^b
\eeq
\beq
\Omega(x,x') = \Omega_0(x,x') + \frac{1}{2} g_{ab,c}(x')(x-x')^a
(x-x')^b(x-x')^c + O^\infty(\Omega_0^2(x,x'))
\eeq
in coordinate neighbourhoods of $(x,x') \in M \times M$, and extended to
all of $M \times M$ as a smooth and positive function. Clearly we have
\beq
G(x,x') = \Omega^{-1/2}(x,x') + \frac{\mu(x')}{2} +
O(\Omega_0^{1/2}(x,x'))
\eeq
\beq
\partial_x [G(x,x') - \Omega^{-1/2}(x,x')] = O(1),
\eeq
which relations also imply that
\beq
G(x,x') = \Omega_0^{-1/2}(x-x') + O(1)
\eeq
\beq
\partial_x [G(x,x') - \Omega_0^{-1/2}(x,x')] = O(1).
\eeq
It follows from the Appendix of Ref. [11] that there exists a unique
positive solution $h$ of (2.15), smooth on $M \setminus \Lambda$,
bounded and with bounded first derivative on $M$. Thus
\beq
h = O(1), \qquad \partial h = O(1) \qquad \mbox{at } \Lambda.
\eeq
We now come to the construction of maximal initial--data sets
$({\bf R}^3,\wt g,\wt p)$ with ${\bf R}^3$ arising as
$M \setminus \Lambda = S^3 \setminus \Lambda$. Let $p_{ab}$ be a
symmetric tensor, smooth on $M \setminus \Lambda$, which is trace-- and
divergence--free (TT--) with respect to $g_{ab}$, where
$\lambda_1(g) > 0$. Assume that
\beq
p_{ab} = O(\Omega_0^{-3/2}), \qquad
\partial p_{ab} = O(\Omega_0^{-2}) \quad \mbox{at } \Lambda.
\eeq
Next solve Equ. (2.13), resp. (2.15) by taking $\rho = p_{ab} p^{ab}$.
It then follows that, with $\wt g_{ab} = \Phi^4 g_{ab}$,
\beq
\R[\wt g] = \frac{1}{8} \wt p_{ab} \wt p^{ab} \qquad \mbox{on }
{\bf R}^3,
\eeq
where $\wt p_{ab} = \Phi^{-2} p_{ab}$. Furthermore
\beq
\wt D_a \wt p^a{}_b = 0, \qquad \wt p_a{}^a = 0.
\eeq
Thus $({\bf R}^3,\wt g,\wt p)$ is a maximal solution of the Einstein
vacuum constraints. Setting
\beq
m = \mu + \left. 2h \right|_\Lambda
\eeq
we have that $(\wt r := \Omega_0^{-1/2})$
\beq
\wt g_{ab} = \left( 1 + \frac{m}{2 \wt r}\right)^4 \delta_{ab} +
O \left( \frac{1}{\wt r^2}\right)
\eeq
\beq
\wt\partial \left[\wt g_{ab} - \left( 1 + \frac{m}{2 \wt r}\right)^4
\delta_{ab}
\right] = O \left(\frac{1}{\wt r^3}\right)
\eeq
\beq
\wt p_{ab} = O \left( \frac{1}{\wt r^3}\right), \qquad
\wt \partial \wt p_{ab} = O \left( \frac{1}{\wt r^4} \right)
\eeq
in coordinates $\wt x^a = \Omega^{-1}(x^a + \frac{1}{2} \Gamma^a_{bc}|_\Lambda
x^b x^c)$. Thus $(\wt g,\wt p)$ is asymptotically flat (in fact:
asymptotically Schwarzschildian) with mass $m$ and vanishing momentum.

We now recall how TT--tensors $p_{ab}$ having the asymptotic behaviour
(2.23) are constructed. This is done on ${\bf R}^3$ with metric $g'_{ab}$
given by
\beq
g'_{ab} = \Omega^{-2} g_{ab},
\eeq
which satisfies
\beq
g'_{ab} = \delta_{ab} + \left(\frac{1}{\wt r^2}\right), \qquad
\wt \partial g'_{ab} = O^\infty \left( \frac{1}{\wt r^3}\right)
\eeq
in the $\wt x^a$--coordinates.

Again, by conformal invariance of the TT--condition we have to solve
\beq
D'_a p'{}^a{}_b = 0, \qquad p_a{}^b = 0
\eeq
with
\beq
p'_{ab} = O^1 \left( \frac{1}{\wt r^3}\right).
\eeq
More specifically we want to find a TT--tensor $p'_{ab}$ with
\beq
p'_{ab} = \frac{C}{\wt r^5} (3\wt x_a \wt x_b - \wt r^2 \delta_{ab}) + O^1
\left( \frac{1}{\wt r^{3+\ve}} \right),
\eeq
where $C$ is a non--zero constant and $\wt x_a = \delta_{ab} \wt x^b$.
Note that (2.34) agrees asymptotically with (1.7), obtained for the
SS metric. Using [14], this is accomplished as follows. Define the
conformal Killing operator $L$, acting on covector fields $\lambda_a$
by
\beq
(L'\lambda)_{ab} = D'_a \lambda_b + D'_b \lambda_a - \frac{2}{3}
g'_{ab}(D'{}^c \lambda_c)
\eeq
and
\beq
(D' \cdot L'\lambda)_b := D'{}^a(L'\lambda)_{ab} = \Delta' \lambda_b +
\frac{1}{3} D'_b (D'{}^a \lambda_a) + \R'{}^b_a \lambda_b.
\eeq
With our asymptotic conditions on $g'$, the operator $D' \cdot L'$
is an isomorphism from the weighted H\"older space
$C^{k+2,\alpha}_\ve({\bf R}^3)$ to $C^{k,\alpha}_{\ve+2}({\bf R}^3)$
for all $k \in {\bf N}$ and all $0 < \ve < 1$ (see [15] for the
precise definitions). We now seek fields
$\st{1}{\lambda}_a$, $\st{2}{\lambda}_a$, $\st{3}{\lambda}_a$,
$\st{4}{\lambda}_a$
solving
\beq
(D' \cdot L' \lambda)_a = 0
\eeq
with
\beqa
\st{1}{\lambda}_a &=& \mu_a + O^\infty \left( \frac{1}{\wt r^\ve}\right) \\
\st{2}{\lambda}_a &=& \mu_{ab} \wt x^b  +
O^\infty \left( \frac{1}{\wt r^\ve}\right) \\
\st{3}{\lambda}_a &=& \ve_{abc} \wt x^b \kappa^c  +
O^\infty \left( \frac{1}{\wt r^\ve}\right) \\
\st{4}{\lambda}_a &=& 6 \wt x_a \nu  +
O^\infty \left( \frac{1}{\wt r^3}\right)
\eeqa
where $\ve_{abc}$ is the Euclidean volume element, $\mu_a$, $\mu_{ab}$,
$\kappa^c$ and $\nu$ are constants with $\mu_{ab}$ symmetric and tracefree
w.r. to $\delta_{ab}$. This is done by observing that the leading terms in
(2.38--41), say $\sigma_a$, have $D' \cdot L' \sigma$ of fast decay,
setting $\lambda_a = \sigma_a + \delta \lambda_a$ and solving
\beq
D' \cdot L' \delta \lambda = - D' \cdot L' \sigma.
\eeq
Denote by $\st{A}{\lambda}_a$ a basis of the 12--dimensional vector space
spanned by
$\st{1}{\lambda}_a$, $\st{2}{\lambda}_a$, $\st{3}{\lambda}_a$,
$\st{4}{\lambda}_a$
and define a tracefree tensor $Q_{ab} = Q_{(ab)}$ by
\beq
Q_{ab} = f \sum_{A=1}^{12} c_A (L' \st{A}{\lambda})_{ab},
\eeq
where $f \in C_0^\infty({\bf R}^3)$, non--negative and not identically
zero. Then we seek constants $c_A$ solving the linear system
\beq
\sum_B E_{AB} c_B = - d_A,
\eeq
where $d_A$ is the vector ($\mu_a = 0$, $\mu_{ab} = 0$, $\kappa^a = 0$,
$\nu = 12 \pi C)$ and
\beq
E_{AB} = E_{(AB)} = \int_{{\bf R}^3} f(L' \st{A}{\lambda})_{ab}
(L' \st{B}{\lambda})^{ab} dV(g').
\eeq
It is shown in [11] that, when $g'_{ab}$ has no conformal isometry,
$E_{AB}$ has trivial null space, and (2.44) has a  unique solution.
If $g'_{ab}$ is only conformally non--flat, $E_{AB}$ has at most a
1--dimensional null space given by
$\bar c_A = (\bar \mu_a, \bar \mu_{ab} = 0, \bar \kappa^c, \bar \nu = 0)$
where $\lambda_a = \st{1}{\lambda}_a(\bar \mu_b) +
\st{2}{\lambda}_a(\bar \kappa^c)$ is a conformal Killing vector.
Clearly $d_A$ is orthogonal to this null space so that (2.44) can
still be solved, and solved uniquely if we require in addition that
\beq
\sum_A \bar c _A c_A = 0.
\eeq
Thus, given $C$ in the definition of $d_A$, we can find $c_A$,
whence $Q_{ab}$. For this $Q_{ab}$ we solve
\beq
(D' \cdot L' W)_a = D'{}^b Q_{ab}.
\eeq
There is a unique solution $W_a$ to any equation of the form
\beq
(D' \cdot L' W)_a = j_a,
\eeq
with $j \in C^{k+2,\alpha}_\ve({\bf R}^3)$, lying in
$C^{k,\alpha}_\ve({\bf R}^3)$. Writing the differential operator as it's
flat--space analogue $\st{o}{D} \cdot \st{o}{L}$ plus the rest, we
infer that
\beq
(\st{o}{D} \cdot \st{o}{L} W)_a = j_a + \delta j_a = \rho_a.
\eeq
Since $g' - \delta = O^\infty(1/\wt r^2)$, the Euclidean components of
$\rho_a$ are again in $C^{k+2,\alpha}_\ve({\bf R}^3)$. Now the arguments
in [11] show that
\beq
W_a = \frac{\wt x_a P_b \wt x^b + 7P_a \wt r^2}{\wt r^3} +
      \frac{\ve_a{}^{bc} \wt x_b L_c}{\wt r^3}
+ \frac{3\wt x_a M_{bc} \wt x^b \wt x^c + 6 M_{ab} \wt x^b \wt r^2}{r^5}
- \frac{C \wt x_a}{2\wt r^3} + \delta W_a,
\eeq
where $\delta W = O^\infty(1/\wt r^{2+\ve})$ with the constants involved in
$\delta W = O(1/\wt r^{2+\ve})$, $\wt\partial \delta W = O(1/\wt r^{3+\ve})$,
$\ldots \wt \partial^k \delta W = O(1/\wt r^{2+k+\ve})$ depending only on
pointwise weighted bounds on $j$ and its first $k$ derivatives.

In the particular case where $j_a = D'{}^b Q_{ab}$, and with the choices
made by (2.45) and by solving (2.46,47), it was shown in [17] that
$P_a$, $L_{ab}$ and $M_{ab}$ are all zero and $C \neq 0$. Thus, by a
computation, the tensor
\beq
p'_{ab} = Q_{ab} - (L' W)_{ab},
\eeq
in addition to being TT, satisfies (2.34).

\section{Critical Sequences}
\renewcommand{\theequation}{\arabic{section}.\arabic{equation}}
\setcounter{equation}{0}

We now consider critical sequences (CS's) $g_n$ of background metrics
on $M$. We require that these metrics all have $\lambda_1(g_n) > 0$
and tend smoothly to a metric $g_\infty$ satisfying
$\lambda_1(g_\infty) = 0$. This requirement of smooth convergence could
be weakened considerably, but we shall not attempt this. It is known
that there are plenty of such sequences. In fact (see [16]), every
metric $g$ with $\lambda_1(g) > 0$ can be deformed into a metric $h$ with
$\lambda_1(h) < 0$ by changing it in an arbitrarily small subset of $M$,
and then one could define $g_t = (1-t)g + th$, which is a continuous
sequence of Riemannian metrics for $t \in [0,1]$. Since, by standard
perturbation theory [17], $\lambda_1(g_t)$ depends continuously on $t$,
there is a $t_0 \in (0,1)$ such that $\lambda_1(g_{t_0}) = 0$ and
$\lambda_1(g_t) > 0$ for $t \in [0,t_0)$. Then define $g_n = (1-t_n)g
+ t_n g_{t_0}$, where $t_n$ is any infinite sequence in $[0,t_0)$ with
$\lim_{n \ra \infty} t_n = t_0$. This is a CS.

It was shown in [10] that, along any CS $g_n$, the time symmetric mass
$\mu_n$ in (2.7) goes to infinity. Furthermore there exists a constant
$E$ independent of $n$ such that
\beq
G_n - \Omega^{-1/2} - \frac{\mu_n}{2} \leq \mu_n E \Omega^{1/2}.
\eeq
(We shall henceforth always denote positive, $n$--independent constants
by $E$ or $E'$ with the understanding that they are not necessarily
identical.) Also
\beq
\partial(G_n - \Omega^{-1/2}) \leq \mu_n E.
\eeq
Note that $\Omega$, which because of Equ.'s (2.4,5) depends on $n$, also has an
upper bound which is independent of $n$. Equ.'s (3.1,2) imply that
\beqa
G_n - \Omega^{-1/2} &\leq& \mu_n E \\
\partial[G_n - \Omega^{-1/2}] &\leq& \mu_n E.
\eeqa
It also follows from [10] that
\beq
G_n - \Omega^{-1/2} \geq \mu_n E,
\eeq
whence
\beq
G_n - \Omega_0^{-1/2} \geq \mu_n E.
\eeq
All constants $E$, $E'$ originate from the Schauder and Harnack inequality
on open subsets of $M$ and thus can be expressed in terms of global
pointwise bounds on $g_n$ and a finite number of it's derivatives.
It follows that these constants, in addition to being independent of $n$,
can be taken to be independent of the choice of source point $\Lambda$.
Thus, in the Green functions $G_n(x,x')$, the same bounds will hold
with $\Omega$ replaced by $\Omega(x,x')$ and in turn by  $\Omega_0(x,x')$, but
with $\mu_n$ now a function of $x'$: $\mu_n = \mu_n(x')$. It is then easily
shown [11] that, in fact, $\mu_n(x')$ has a uniform bound from above and below
in
terms of $\mu_n$ at any other point of $M$, say $\Lambda$. Writing now
$\mu_n$ for $\left. \mu_n \right|_\Lambda$, we get
\beqa
E'[\Omega_0^{-1/2}(x,x') + \mu_n] \leq G_n(x,x') &\leq&
E[\Omega_0^{-1/2}(x,x') + \mu_n] \\
\partial_x[G_n(x,x') - \Omega_0^{-1/2}(x,x')] &\leq& E \mu_n.
\eeqa
Let $N_\Lambda$ be a fixed coordinate neighbourhood of $\Lambda$ with
chart $y^a$ centered at $\Lambda$. Let $g_\delta$ be a metric on $M$
which coincides with the flat metric $\delta_{ab}$ in $N_\Lambda$ and
let $\Omega_\delta^{1/2}(x,x')$ be the Euclidean $|y-y'|$, extended
smoothly as a positive function to all of $M \times M$. Clearly we
have
\beq
E' \Omega_\delta(x,x') \leq \Omega_0(x,x') \leq E \Omega_\delta(x,x').
\eeq
Thus
\beq
E'[\Omega_\delta^{-1/2}(x,x') + \mu_n] \leq G_n(x,x') \leq
E[\Omega_\delta^{-1/2}(x,x') + \mu_n]
\eeq
\beq
\partial_x G_n(x,x') \leq D[\Omega_\delta^{-1}(x,x') + \mu_n].
\eeq
Given the CS $g_n$ we can decompactify as shown in the last section,
thus obtaining a sequence $g'_n$ of asymptotically flat metrics with
components converging to a metric $g'_\infty$ in the H\"older norms
$C_\ve^{k+3,\alpha}({\bf R}^3)$ for all $k \in {\bf N}$.
For each $g'_n$ and for $g'_\infty$ there is the operator
$D' \cdot L'$, written down in Equ. (2.36) which maps
$C^{k+2,\alpha}({\bf R}^3)$ isomorphically onto
$C^{k,\alpha}_\ve ({\bf R}^3)$. Thus, if there is a sequence of sources
$j_n$ converging to some $j_\infty$ in the $C^{k,\alpha}_\ve$--norm,
the solution $W_n$ of $(D'_n \cdot L'_n)W_n = j_n$ converges to
$W_\infty$ in the $C^{k+2,\alpha}_\ve$--norm. Hence the fields
$\st{A}{\lambda}_n$, defined in the last section, converge to
$\st{A}{\lambda}_\infty$ corresponding to the metric $g'_\infty$.
In particular the constants involved in the remainder terms in
(2.38 -- 2.41) can be taken to be independent of $n$.

Let us now, for ease of presentation, suppose that the sequence $g_n$
and it's limit $g_\infty$ are ``generic''. By this we mean that no linear
combination of the $\st{A}{\lambda}$'s for these metrics is a
conformal Killing vector. Then, by performing for each $n$ the
procedure described between Equ. (2.43) and Equ. (2.51), we obtain
fields $W_n$  with $P_a = L_a = M_{ab} = 0$ and fixed $C \neq 0$
converging, in the $C_\ve^{k+2,\alpha}$--norms, to a $W_\infty$
with the same properties.  Splitting the operators
$D'_n \cdot L'_n$ into a flat--space part and the rest, it is standard
to see  [18] that the bounds on
$\delta W$ in (2.50) are uniform in $n$. When we now undo the
decompactification we obtain, for each metric $g_n$ in the CS, a
TT--tensor $p_{ab}$ on $M$ obeying
\beq
p_{ab} = C \frac{3 D_a(\Omega^{1/2}_0) D_b(\Omega_0^{1/2}) -
\left. g_{ab}\right|_\Lambda}{\Omega_0^{3/2}} + O^\infty
(\Omega_0^{-(3+\ve)/2}), \qquad C > 0
\eeq
with $O^\infty$ independent of $n$. We now have to solve, for each
$n$, the equation (2.13) with $\rho = p_{ab} p^{ab}$.
This, in turn, requires to solve the integral form of equation (2.15),
namely
\beq
h(x) = \frac{1}{8} \int_M G(x,x') \frac{G^{-7}(x') \rho(x')}
{[1 + G^{-1}(x') h(x')]^7} dV_g(x').
\eeq
Suppose $x \in N_\Lambda$. Then the integral in (3.13) is dominated by
the contribution coming from integration over $x' \in N_\Lambda$.
Using (3.9, 10, 11, 12) and that
\beq
E' \Omega_\delta^{7/2}(x') \leq G^{-7}(x'),
\eeq
we see, that $h$ is bounded by positive constants times the sum of the
following integrals
\beq
A(y) = \int_{|y'| \leq R} \frac{1}{|y-y'|} \left( \frac{1}{|y'|} +
\mu_n D \right)^{-7} \frac{1}{|y'|^6} d^3y'
\eeq
and
\beq
B(y) = \mu_n \int_{|y'| \leq R} \left( \frac{1}{|y'|} + \mu_n D\right)^{-7}
\frac{1}{|y'|^6} d^3y',
\eeq
where $R$ is a bound on the size of $N_\Lambda$. It is elementary to check
that
\beq
A = O(\mu_n^{-3}), \qquad B = O(\mu_n^{-3})
\eeq
with $O$ independent of $n$.

Similarly we find that $\partial h$ is uniformly bounded by
\beq
C(y) = \int_{|y'| \leq R} \frac{1}{|y-y'|^2}
\left( \frac{1}{|y'|} + \mu_n D\right)^{-7} \frac{1}{|y'|^6} d^3y'
\eeq
and
\beq
D(y) = \mu_n \int_{|y'| \leq R} \left( \frac{1}{|y'|} +
\mu_n D\right)^{-7} \frac{1}{|y'|^6} d^3 y'.
\eeq
We find that
\beq
C(y) = O(\mu_n^{-3}), \qquad D(y) = O(\mu_n^{-2}).
\eeq
Consequently
\beq
h = O(\mu_n^{-3}), \qquad \partial h = O(\mu_n^{-2}).
\eeq
The above estimates provide information on the quantities
$\Theta_\pm$ associated with surfaces ``$\Omega$ = small constant''
as $n$ gets large. Recall that
\beq
\Theta_\pm = H \mp \wt p_{ab} \wt n^a \wt n^b.
\eeq
The quantity $H$, in the present case, is given by
\beq
H = \wt D_a \wt n^a = - \frac{4 \Omega^{-1/2} \Phi^{-3}}{(\Omega_c
\Omega^c)^{1/2}} \left[ \frac{1}{4} \Omega^{1/2} \Phi
\left( g^{ab} - \frac{\Omega^a \Omega^b}{\Omega_d \Omega^d}\right)
\Omega_{ab} + \Omega^{1/2} \Phi_a \Omega^a \right].
\eeq
In $N_\Lambda$ we can, for each $n$, change the $y$--coordinates
centered at $\Lambda$ to $\bar y^a$ for which
$\left. g_{ab}\right|_\Lambda = \delta_{ab}$ and then change these,
in $N_\Lambda \setminus \Lambda$ to $(\theta,\vp,r)$, where
$(\theta,\vp,r)$ is related to $\bar y^a$ like standard spherical
coordinates to standard Euclidean coordinates. The quantity
$$
b = \Delta \Omega - \frac{\Omega^a \Omega^b g_{ab}}{\Omega_c \Omega^c}
$$
appearing in (3.23) can be expanded as
\beq
b = 4 + f_1 r + f_2 r^2 + f_3 r^3 + O(r^4)
\eeq
with $f_1$, $f_2$, $f_3$ smooth functions on $(\theta,\vp) \in S^2$.

Recall also that $\Phi = G + h$, $G$ being the conformal factor giving the
time--symmetric solutions conformal to the given background metrics $g_n$.
The quantity $G$, in turn, [19] can be written as a sum
\beq
G = G^{loc} + F,
\eeq
where $G^{loc}$ is the Hadamard fundamental solution of $L_g$ in a
neighbourhood of $\Lambda$, whence $F$ is smooth and satisfies
$L_g F = 0$ in this region. By [19] this region can be chosen independent
of $n$. It is known [19] that $G^{loc}$ can be written as follows
\beq
G^{loc} = \Omega^{-1/2} V + W,
\eeq
where $V = 1 + O(\Omega)$, $\left. W\right|_\Lambda = 0$ and each of
$V$ and $W$ permit an expansion in terms of $s^2$, the (geodesic
distance)$^2$ to the point $\Lambda$. Clearly $s^2$ can be expanded as
$s^2 = r^2 + g_1 r^3 + \ldots$ with $g_1,\ldots$ smooth on $S^2$, and
the same is true for $\Omega$. Similar expansions hold for
$\Omega_a$ and $G^{loc}$. Making $N_\Lambda$ smaller, if necessary,
inserting this into (3.23) and using (3.21) we find that the square
bracket in (3.23) can be written as $X + Y$ with
\beq
X = 1 - \frac{\mu_n}{2} r + \mu_n a_1 r^2 + \mu_n a_2 r^3 +
O(\mu_n r^4) + O(\mu_n^{-3} r) + O(\mu_n^{-2} r^2).
\eeq
The coefficients $a_1$ and $a_2$ are smooth on $S^2$ and remain bounded as
$n \ra \infty$. We now set
\beq
r_n = \frac{2}{\mu_n} + \frac{8a_1}{\mu_n^2} + \frac{16a_2}{\mu_n^3} +
\frac{64a_1^2}{\mu_n^3}
\eeq
and take $n$ large enough so that $(\theta,\vp,r_n) \in N_\Lambda$.
Then Equ. (3.28) defines an embedded 2--sphere $\Sigma_n \subset
N_\Lambda$. Evaluating $X$ along $\Sigma_n$, a computation gives that
\beq
\left. X\right|_{\Sigma_n} = (\mu_n^{-3}).
\eeq
Thus the surface $\Sigma_n$ becomes more and more minimal as
$n \ra \infty$. The second term in Equ. (3.21) can be estimated from
\beq
\wt p_{ab} \wt n^a \wt n^b = \frac{4 \Omega^{-1/2} \Phi^{-3}}
{(\Omega_c \Omega^c)^{1/2}} \left[ \frac{1}{4} (\Omega_d \Omega^d)^{1/2}
\Omega^{1/2} \Phi^{-3} p_{ab} n^a n^b\right].
\eeq
Now insert (3.12) into (3.30), using that
\beq
r = \Omega_0^{1/2} + O(\Omega_0) = \Omega_0^{1/2} + O(\mu_n^{-2}), \qquad
D_a r = D_a \Omega_0 + O(\mu_n^{-1}), \qquad
D_a r D^a r = 1 + O(\mu_n^{-1}).
\eeq
We find that the square bracket in Equ. (3.30) is equal to
\beq
Y = \frac{1}{4} \Phi^{-3} [3\mu_n C + O(\mu_n^{1-\ve})].
\eeq
But, from (3.1),
\beq
\left. \Phi\right|_{\Sigma_n} = \frac{3\mu_n}{2} + O(1),
\eeq
so that
\beq
Y = \frac{2C}{9} \mu_n^{-2} + O(\mu_n^{-2-\ve}).
\eeq
Thus, as $n$ gets large,
\beq
\Theta_+ < 0, \qquad \Theta_- > 0 \qquad \mbox{on } \Sigma_n.
\eeq
We have thus proved the
\paragraph{Theorem:} Let $g_n$ be a generic initial sequence and
$(\wt g_n,\wt p_n)$ the unique initial--data set constructed for each
$n$ from the ``free data'' $g_n$, $p_n$ with $p_n$ coming from the
``same tensors'' $Q_{ab}^n$, so that $p_n$ is of the form (3.12) with
a fixed constant $C > 0$. Then the 2--spheres $\Sigma_n$ given by (3.28)
are future trapped for sufficiently large $n$.

We call a CS special when, for each $n$ and for $g_\infty$, there is a
linear combination of the $\lambda$--fields defined in Sect.~2, which is
a conformal Killing vector. (This has to be unique for $g_\infty$ and for
$g_n$ with $n$ large.) Then the above reasoning goes through unchanged,
except that Equ. (2.44) is now solved with the side condition (2.46).
Hence the above Theorem remains true with ``generic CS'' changed into
``special CS''.

The construction we have described for given $g_n$ and given function $f$
appearing in (2.43), gives rise to a unique sequence of initial--data
sets. There is of course much more freedom in the choice of extrinsic
curvature. It is known [20] that, on the compact manifold $M$, there is an
infinite--dimensional set of smooth TT--tensor $\tau_{ab}$ for any
background metric $g$. Adding such a tensor to the one constructed
from $Q_{ab}$ only changes $\wt p_{ab}$ at higher multipole order and
thus gives rise to a critical sequence for which the Theorem remains to
be true.

Finally, we could generalize our result by including angular momentum, that
is to say construct data for which the $W$--field of Equ. (2.50) has no
linear--momentum term $P_a$ and no quadrupole--contribution $M_{ab}$,
but both $L_a$ and $C$ non--zero. The reason is that the angular momentum
term in $\wt p_{ab}$ gives, to leading order, a vanishing contribution
to $\wt p_{ab} \wt n^a \wt n^b$, so that $\wt p_{ab} \wt n^a \wt n^b$
will asymptotically still have a sign for large $n$.

\end{document}